! Tex program = xelatex
\documentclass[]{raa} 

\usepackage{graphicx,times}             
\usepackage{natbib}
\usepackage{amssymb,amsmath}
\bibpunct{(}{)}{;}{a}{}{,}

\begin{document}

  \title{The Stellar Abundances and Galactic Evolution Survey (SAGES) III}
 \subtitle{The g/r/i-band Data Release}

   \volnopage{Vol.0 (20xx) No.0, 000--000}      
   \setcounter{page}{1}          

   \author{Chun Li
      \inst{1}
   \and Zhou Fan
      \inst{1,2}
    \and Gang Zhao
      \inst{1,2}
   \and Wei Wang
      \inst{1}
      \and Jie Zheng
      \inst{1}
         \and Kefeng Tan
      \inst{1}
         \and Jingkun Zhao
      \inst{1}
      \and Yang Huang
      \inst{1}
         \and Haibo Yuan
      \inst{2}
         \and Kai Xiao
      \inst{1,2}
         \and Yuqin Chen
      \inst{1}
         \and Haining Li
      \inst{1}
         \and Yujuan Liu
      \inst{1}
         \and Nan Song
      \inst{1,3}
         \and Ali Esamdin
      \inst{4}
         \and Hu-Biao Niu
      \inst{4}
         \and Jin-Zhong Liu
      \inst{4}
         \and Guo-Jie Feng
      \inst{4}
   }

   \institute{CAS Key Laboratory of Optical Astronomy, National Astronomical Observatories, Chinese Academy of Sciences, Beijing 100101, P.R.China; {\it lichun@nao.cas.cn, zfan@nao.cas.cn}\\
        \and
             School of Astronomy and Space Science, University of Chinese Academy of Sciences, Beijing 100049, China\\
        \and
             China Science and Technology Museum, Beijing 100101, P.R.China\\
        \and Xinjiang Astronomical Observatory, Urumqi, Xinjiang 830011, P.R. China\\
\vs\no
   {\small Received 20xx month day; accepted 20xx month day}}

\abstract{The Stellar Abundances and Galactic Evolution Survey (SAGES) is a multi-band survey that covers the northern sky area of $\sim$12000 deg2. Nanshan One-meter Wide-field Telescope (NOWT) of Xinjiang Astronomical Observatory (XAO) carried out observations on g/r/i bands. We present here the survey strategy, data processing, catalog construction, and database schema. The observations of NOWT started in 2016 August and was completed in 2018 January, total 17827 frames were obtained and $\sim$4600 deg2 sky areas were covered. In this paper, we released the catalog of the data in the g/r/i bands observed with NOWT. In total, there are 109,197,578 items of the source records. The catalog is the supplement for the SDSS for the bright end, and the combination of our catalog and these catalogs could be helpful for source selections for other surveys and the Milky Way sciences, e.g., white dwarf candidates and stellar flares.
}

   \authorrunning{ }            
   \titlerunning{ }  

   \maketitle

%
%
\section{Introduction}           
\label{sect:intro}
Astronomy is a science based on observation. Large field sky surveys are of great significance to the study of the structure and evolution of the Galaxy. In recent decades, massive discoveries have been made base on a lot of sky survey projects such as Gaia, LAMOST, GSC, 2MASS, etc. SkyMapper is a southern sky survey that is efficient at identifying metal-poor stars. Hence we launch a similar project, SAGES for the northern sky area.

The Stellar Abundances and Galactic Evolution Survey (SAGES) is a multi-band survey that covers the northern sky. Nanshan One-meter Wide-field Telescope (NOWT) of Xinjiang Astronomical Observatory (XAO), Chinese Academy of Sciences (CAS) carried out observations on g/r/i passbands. We present here the survey strategy, data processing, catalog construction, and database schema. The observation of NOWT started in 2016 August and was completed in 2018 January, total 17827 frames were obtained and $\sim$4600 deg2 sky areas were covered.

SAGES is a northern sky photometry survey with eight passbands, that intends mainly to search for metal-poor stars in the Galaxy\citep{sages23}. The survey aims to derive stellar atmospheric parameters for a few hundred million stars. The SAGES photometric system is unique and defined by the SAGES team, which is composed of eight bands(the $u_s$/$v_s$/g/r/i/$\alpha_n$/$\alpha_w$/DDO51), including narrow-band, intermediate bands, and broad bands as well. The sky coverage of the project covers the northern sky region of Dec $ \delta >$ -5$^{\circ}$  and avoids the Galactic disk region of $-10^{\circ} < b < +10^{\circ}$  while avoiding saturation and image contamination that may result from excessive bright stars.  The designed survey area was larger than 12,000 deg2($\sim60\%$ of the northern sky). Since the beginning of operation in 2016, data has been obtained for 5 bands. Among them, the $u_s$ and $v_s$ bands catalog has been released in  Fan et al. (2023). The g/r/i bands were observed by NOWT, data of 4254 sky areas  was covered, and 17827 frames were obtained. After datareduction, photometry, calibration and data combination, 51,149,452 sources were detected. The complete magnitudes for g/r/i passbands are 19.2mag, 19.1mag and 18.2mag, respectively. This part of data has been released and availbable on China-VO platform.

This paper briefly introduces the observation data of Nanshan g/r/i bands in SAGES. The second section will introduce the basic situation of the project. The observation detail will be shown in section 3. The data reduction processing will be described in section 4. In section 5 we show the data quality, and the discussion and prospect in section 6.

\section{Nanshan DATA of SAGES Photometric System}
\label{sect:Obs}
\subsection{The Photometric System of SAGES}

\begin{figure}
    \centering
    \includegraphics[width=12cm]{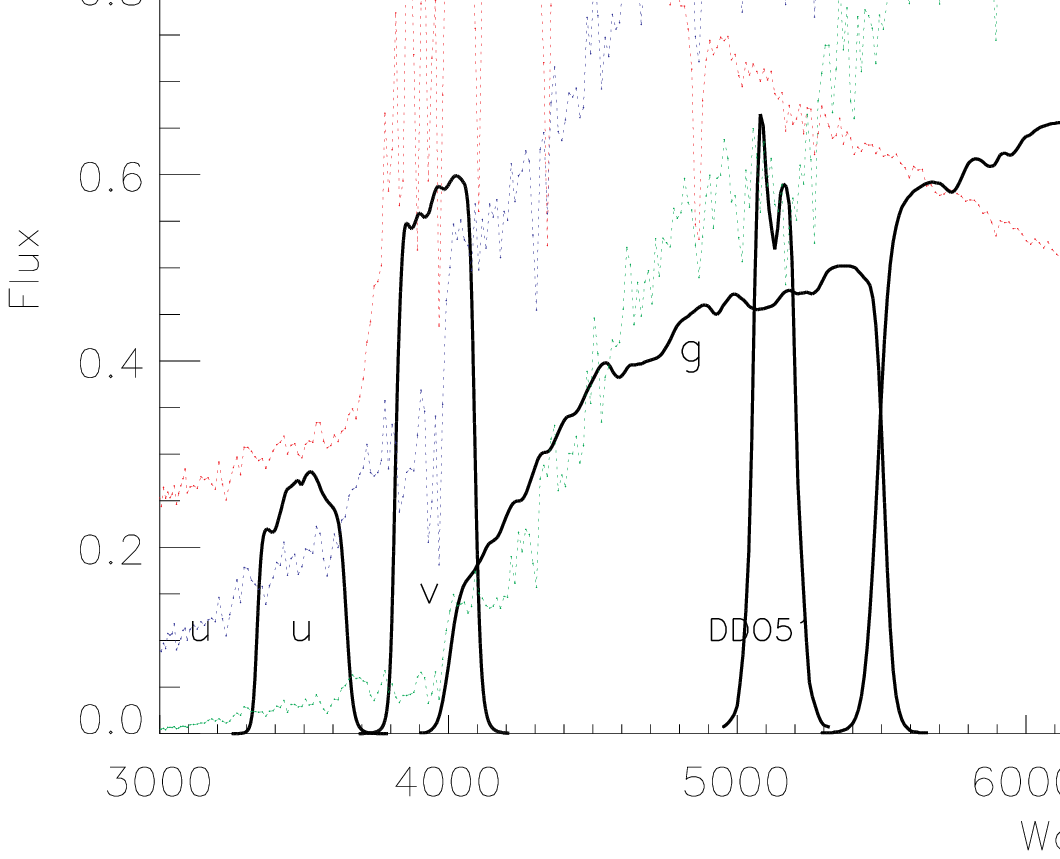}
    \caption{The filter response functions of SAGES photometric system, overplotted with the spectra of typical F-stars (red) G-stars (blue) K-stars (green).}
    \label{fig1}
\end{figure}

Figure~\ref{fig1} and Table~\ref{tab1} present the details of the 8 passbands of SAGES. The $u_s$ is similar to the one in the Str\"{o}mgren-Crawford (SC) system which covers the Balmer jump, and $v_s$ is a self-developed filter that covers the Ca II H\&K absorption lines, which are very sensitive to the metallicity of stars. Therefore, the two passbands have a good identification function for metal-poor stars. The other three bands, g/r/i are the same as SDSS passbands, which are useful in estimating the effective temperature $T_{\rm eff}$ of stars. The observation plan made for g/r/i in SAGES mainly intends to complete the coverage of the whole north sky with the combination of the SDSS catalog. The DDO51 band is sensitive to the gravity of late-type stars. Another two bands, $\alpha_n$ and $\alpha_w$, are used to estimate the interstellar extinctions, as the value of $\alpha_w-\alpha_n$ is only sensitive to the effective temperature $\rm T_{eff}$, and it is independent of interstellar extinction. Thus the effective temperature $\rm T_{eff}$ is easy to be constrained more accurately. Then the difference between them and $\rm T_{eff}$ from other colors carried constrains for  extinction. DDO51 measures the MgH feature in KM dwarfs\citep{bess05}. For further information about SAGES please refer to Fan et al. (2023).

\begin{table}[]
    \centering
        \caption{SAGES photometry system}
    \begin{tabular}{|c|c|c|c|c|c|c|c|c|}
\hline
Band&$u_s$&$v_s$&g&r&i&DDO51&$\alpha_w$&$\alpha_n$\\\hline
wavelength(\AA)&3520&3950&4639&6122&7439&5132&6563&6563\\\hline
width(\AA)&314&290&1280&1150&1230&162&29&136\\\hline
    \end{tabular}
    \label{tab1}
\end{table}

\subsection{The overall observation progress of the SAGES project}
\label{sect:Obs2}
The observation of eight bands of SAGES is processed in different telescopes and sites.

The $u_s$- and $v_s$-bands are observed in Kitt Peak National Observatory(KPNO) of Arizona, USA, with the 90-inch Bok telescope.

Because of the lack of observation time from February to August and weather conditions, the sky area of 12 $< \alpha < 18$ h of R.A is not included in the survey. The u/v passband data has been released in \citep{sages23}.

The observations of g/r/i passbands have been carried out in NOWT already since 2016 August, and ended in 2018 January, which part of the data is introduced in this paper.

From the plan of DDO51, $\alpha_w$ and $\alpha_n$ passbands, multiple telescopes were intended for use, including 1-m Zeiss telescope of Maidanak astronomical observatory (MAO, 66$^\circ$53$^\prime$47$^{\prime\prime}$E, 38$^\circ$40$^\prime$22$^{\prime\prime}$N), which belongs to Ulugh Beg Astronomical Institute (UBAI), Uzbekistan\citep{kf76}, and Xuyi 1-m Schmidt telescope of Purple Mountain Observatory (PMO) of CAS.

Observation of the DDO51 passband with the telescope of NOWT is still going on. The observations of the whole sky coverage of SAGES will be completed in 3 years(2023 to 2026).

\subsection{Observatory and Telescope}
\label{sect:data}

The Nanshan One-meter Wide-field Telescope (NOWT) belongs to Xinjiang Astronomical Observatory (XAO), Chinese Academy of Sciences (CAS), which is located at Nanshan Observatory ($87^{o}$ 10$^\prime$30$^{\prime\prime}$E, $43^{o}$ 28$^\prime$25$^{\prime\prime}$N), which is $\sim$75~km away from Urumqi city, with an altitude of 2088 meters.\citep{bai} The observable night of Nanshan Station is more than 300 days per year, and the clear night is greater than 210. The distribution peak of seeing values around $1.67^{\prime\prime}$, and 80\% of the value obtained at night are below 2.2. At zenith, the sky brightness is around 21.7mag $\rm arcsec^{-2}$ in the V -band.\citep{bai}

NOWT is a one-meter telescope with an Alt-Az mount, operating at prime focus with a field corrector. The parabolic primary mirror's effective diameter is 1 meter and the focal length is $2159 \pm 20mm$\citep{bai}. Pointing accuracy is better than $3^{\prime\prime}$(RMS-error), tacking accuracy is $1.8^{\prime\prime}$ RMS in 60min.\citep{bai}

 Totally $80\%$ of the collected energy is concentrated into a circle with a diameter of less than $1.15^{\prime\prime}$ on a field diameter of 2.4$^\circ$. The pointing accuracy for this telescope is better than $5^{\prime\prime}$ RMS for each axis after pointing model correction \citep{liu13}\citep{bai}.

 The NOWT provides excellent optical quality in photometry which became a great benefit for SAGES.

\subsection{Detectors of Nanshan g/r/i Observation}

The filters of g/r/i passbands used NOWT for SAGES were made in Custom Scientific, Inc, USA, which are the standard SDSS system. 

\begin{figure}
    \centering
   \includegraphics[width=10cm]{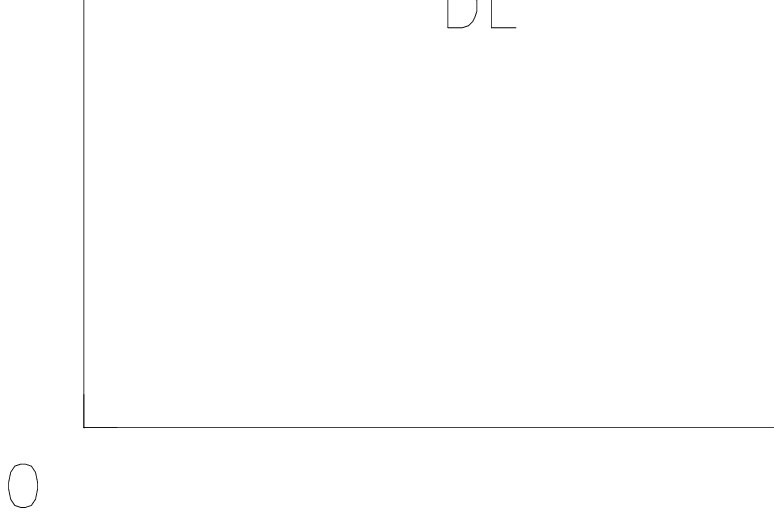}
    \caption{The Andor 4k $\times$ 4k NOWT CCD array, shows 4 amplifiers. The right side shows an overscan area.}
    \label{fig2}
\end{figure}

The CCD camera installed to NOWT was $4096\times 4136$  pixel, designed and integrated by the CCD laboratory of NAOC\citep{bai}. The camera was placed at the prime focus of the telescope and cooled by liquid nitrogen. The camera has 4 amplifiers for fast readout which is shown in Figure~\ref{fig2}. The physical size of the pixel was $12\times 12 \mu m$, with given the field of view $1.3^\circ\times1.3^\circ$, and the pixel scale of $1.124^{\prime\prime}$\citep{bai}.

For the overscan, as we can see in Figure~\ref{fig2}, 32-pixel columns in total are set under scan pixels for each amplifier, which is used to correct image data for bias.

\subsection{Observations}
The planned survey area was given in section 1. The g/r/i bands of the mission are mainly to complete the sky area of the SDSS survey and SAGES. Because of the latitude of NOWT and telescope location, the sky areas only for Dec $>5^{\circ}$ were observed.

\begin{figure}
    \centering
        \includegraphics[width=10cm]{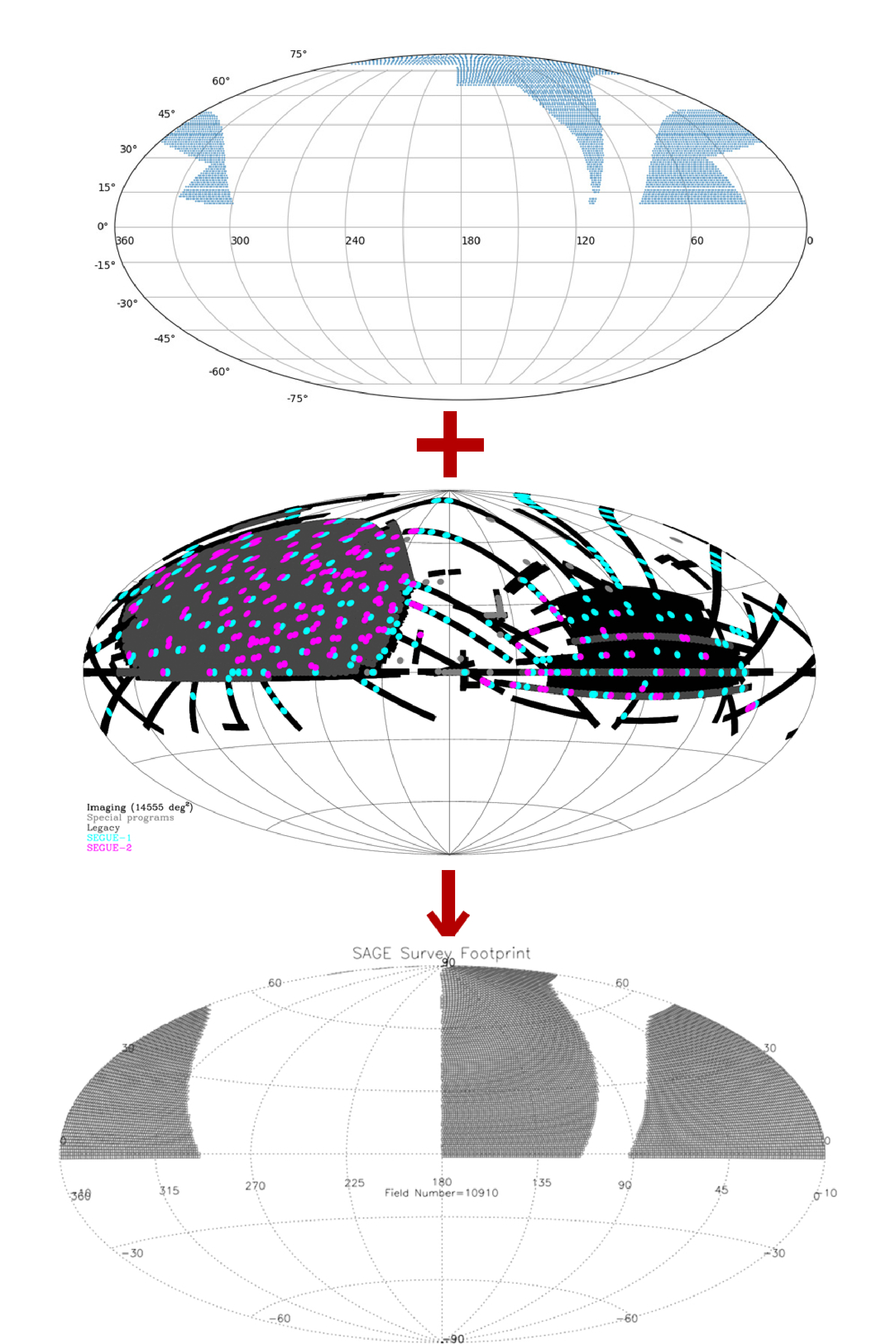}
    \caption{The sky coverage for SAGES in g/r/i passband. The designed area can be combined with the SDSS data for the completeness of the catalog, including 4254 blocks. The upper panel present the designed area of g/r/i passbands, the middle panel shows the SDSS Legacy Sky Coverage(https://www.sdss4.org/dr17/), the bottom panel is the sky coverage of the whole SAGES project.}
    \label{fig3}
\end{figure}

\subsection{Observing Strategy and Coverage}

The field of view of NOWT is $\sim 1{^{\circ}}.3\times1^{\circ}.3$, so we have designed each survey field as $1^{\circ}.04 \times 1^{\circ}.04$. Thus the designed $\sim$4600 deg2 sky area were devided into 4254 blocks for observation. The fields on the same Dec are called a stripe, and the adjacent 5 fields are combined as a survey block, so we will reduce the long moving between blocks. The overlap area between neighbor fields will help in the flux calibration and ensure the coverage of the whole sky. We named each field with an integer sequence. Since SDSS had covered a large area of the southern sky, we need only to finish the rest part. The blocks of NOWT are displayed in Figure ~\ref{fig3}.

We have developed a strategy pipeline that helps to generate observation scripts for each night. The pipeline works in these ways: ensuring no duplicated observation of each field, choosing proper blocks according to the observation date, excluding the fields near the moon, and sorting the blocks to make sure airmass will be minimum. The pipeline will also provide the animated progress of observations so we can check our observation progress. If the observation was interrupted by bad weather or other opportunity targets, the report of the pipeline will help the operators to resume from a proper position.

\section{Data Reduction}

\subsection{Image Correction}

One pipeline was built for the photometry and astrometry steps by our team. The data correction procedure includes the overscan correction, biases combination, flats combination, and survey images correction. Twilight flats are used for correction in NOWT runs.

For overscan, the median value of each row was subtracted. For bias, we took a set of 10 bias images before and after observation, and use the median of bias for correction. Then all survey images and flat-field images process bias removal.

For flat, a set of 10 skylight flat frames for each band was taken before and after observation, if weather allowed. After removing of bias, the flat images were normalized and combined with median values.

The survey images are executed through overscan correction, bias removal, and flat correction procedure.

The number of images after correction are 5283 frames in the g band, 4849 frames in the r band , and 5314 frames in the i band.

The tool Source Extractor (SE)\citep{ber96} was used to detect the sources and perform the photometry. Astrometry tool Scamp (Software for Calibrating AstroMetry and Photometry)\citep{ber06} was used for the astrometric procedure. Both of SE and SCAMP are well-developed and widely used in data reduction of photometry surveys. The Position and Proper Motion Extended (PPMX)\citep{roser} catalog is adopted as the astrometric reference for its relatively good accuarcy\citep{zheng18,zheng19}.

We run SE for the corrected images using the position information in the fits head to produce a preliminary source catalog Then we run SCAMP twice with different parameters to get more precise astrometry results. With the WCS information, run SE again to obtain a catalog of instrumental magnitude. The output parameter Mag\_Auto of SE was used as the primary instrumental magnitude. 

The data fields of the final local catalog of each image are listed in Table~\ref{tab2}. The FLAG\_G/R/I are the standard SE flags. For detail of flags please refer to \citet{sages23} and \citet{ber96}.

\begin{table}[]
    \centering
        \caption{Parameters of Output catalog}
    \begin{tabular}{|l|l|}
    \hline
SAGE\_ID&ID for SAGES \\ \hline
RA&Right ascension of object (J2000)\\\hline
DED&Declination of the object (J2000)\\\hline
MAG\_G/R/I&Magnitude in the g/r/i passband\\\hline
ERR\_G/R/I&Error of magnitude in g/r/i passband\\\hline
OBS\_TIME\_G/R/I&Time of observation of certain frame\\\hline
FLAG\_G/R/I&Flag from Source Extractor of certain frame\\\hline
    \end{tabular}

    \label{tab2}
\end{table}

\subsection{Flux Calibration}

\citet{xiao2023a} utilized both the spectroscopy-based Stellar Color Regression method (SCR method; \citep{yuan15}) and the photometric-based SCR method (SCR$^\prime$ method) to construct a total of approximately 2.6 million dwarfs as standard stars, achieving an accuracy of about 0.01--0.02\,mag per band. They then performed relative photometric calibration of the NOWT's g/r/i bands imaging data. Absolute calibration was conducted using the corrected Pan-STARRS DR1 (PS1) photometry by \citet{Xiao and Yuan(2022)} and \citet{xiao2023b}, establishing the transformation relationship between the calibrated magnitudes from the Nanshan One-meter Wide-field Telescope and the corrected PS1 magnitudes. A comprehensive analysis of repeated sources in adjacent images revealed a remarkable internal consistency of approximately 1--2\,mmag across all filters. By employing the synthetic photometry method with Gaia DR3 BP/RP spectra, the PS1 magnitudes were synthesized, resulting in a  photometric calibration uniformity of approximately 2.4\,mmag, 2.3\,mmag, and 0.9\,mmag photometric calibration uniformity for the g/r/i bands, respectively, within each field with diameter of 1.3 degrees. During the photometric calibration process, the dependence of the CCD gain and stellar flat on the observation time is discussed, as well as the CCD position-dependent residual of stellar flat correction. More details can be found in \citet{xiao2023a} and \citet{sages23}.

\section{Data Quality}
In the Nanshan part of the data, the count of g passband images involved in calibration is 5096, and 4551 of r passband, and 5153 of i passband. The total count of survey images is 14800.

The total sources detected are 31,894,325 in g band, 38,711,424 in r band, and 38,531,820 in i band. After a combination of all sources, we have 51,149,452 sources in total. Among them, 21,819,140 were detected in all g/r/i passbands.

\subsection{Astrometry precision}

\begin{figure}
    \centering
    \includegraphics[width=10cm]{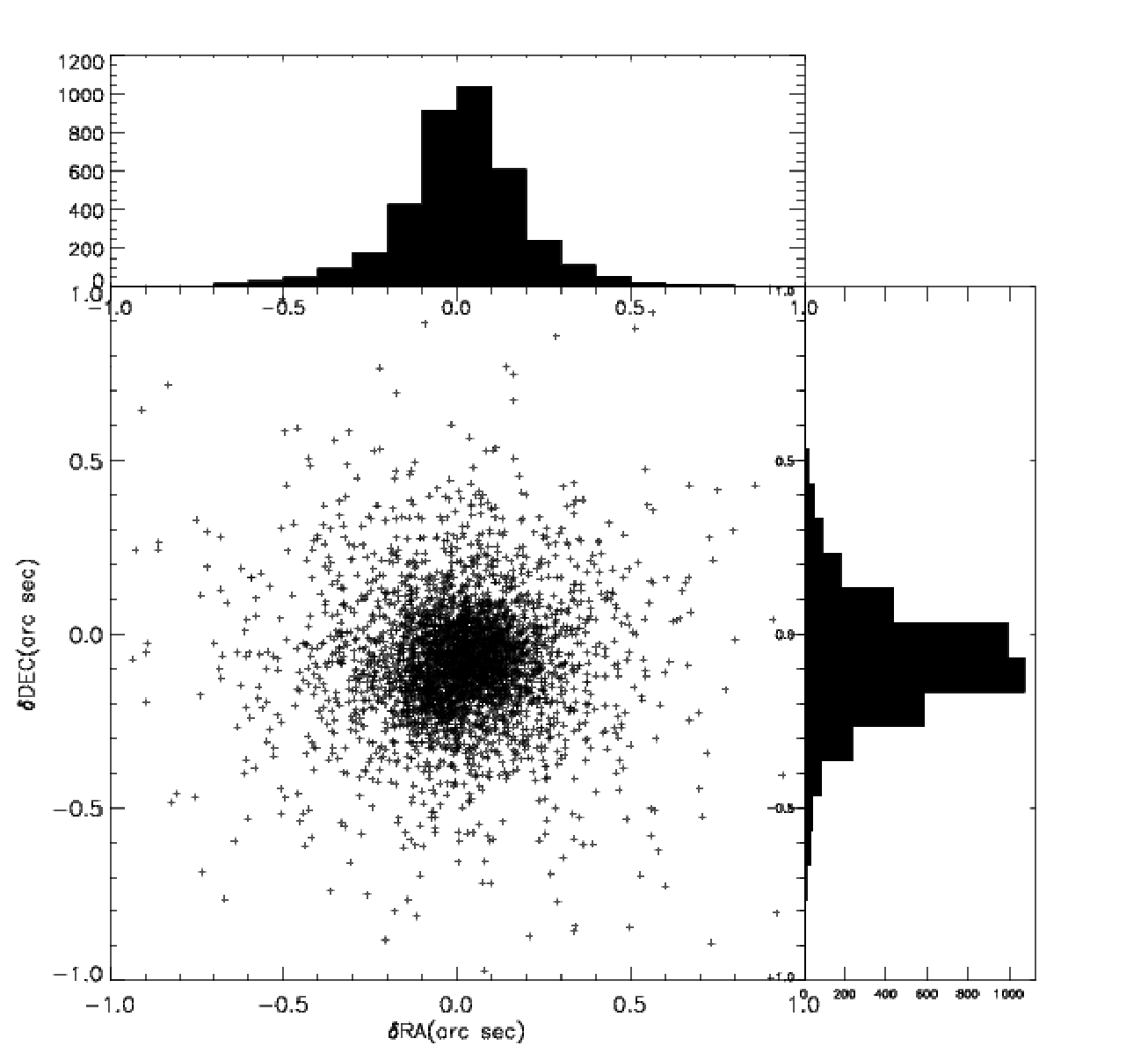}
    \caption{The precision of astrometry for Nanshan run. The external astrometric errors are estimated by comparing the difference between the coordinates of one typical image (named n8037.0172 of g band) and those from reference catalog PS1.}
    \label{fig4}
\end{figure}

Figure~\ref{fig4} presents the typical astrometric error. The data is from one frame named n8037.0172 in g band. The external astrometric errors are estimated by comparing the difference between the coordinates of one typical image and those from reference catalog PS1. The rms of difference between the coordinates(caculated from all detected sources) are 0.28$^{\prime\prime}$ for g band, 0.29$^{\prime\prime}$ for r band and 0.30$^{\prime\prime}$ for i band, which are good enough for the WCS and astrometry.

\subsection{The Limiting Magnitudes}

\begin{figure}
    \centering
    \includegraphics[width=12cm]{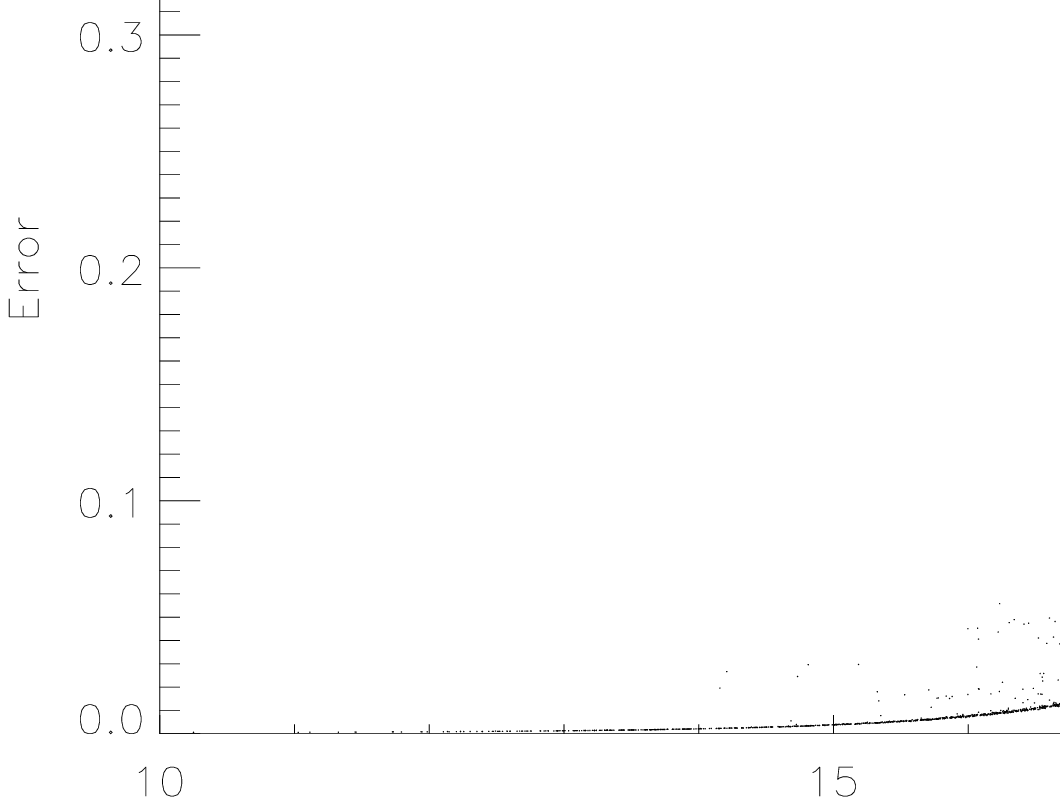}
    \caption{The photometric uncertainties (from Poisson statistics) vs. magnitude in Nanshan run for one typical image named n7634.0303. The limiting magnitude is $\sim$ 16.5 mag, with uncertainties of 0.01 mag, which corresponds to S/N $\sim$ 100.}
    \label{fig5}
\end{figure}

Figure~\ref{fig5} shows a typical magnitude vs photometric error diagram in r band. The photometric error is from Poisson statistics, not including the calibration errors. We can see that for the limiting magnitude of S/N$\sim10$(uncertainty of 0.1)  is $\sim$19.0. The scatter is due to the local background fluctuation which may be caused by contamination of nearby sources and nonuniformity of skylight background.

\begin{figure}
    \centering
   \includegraphics[width=10cm]{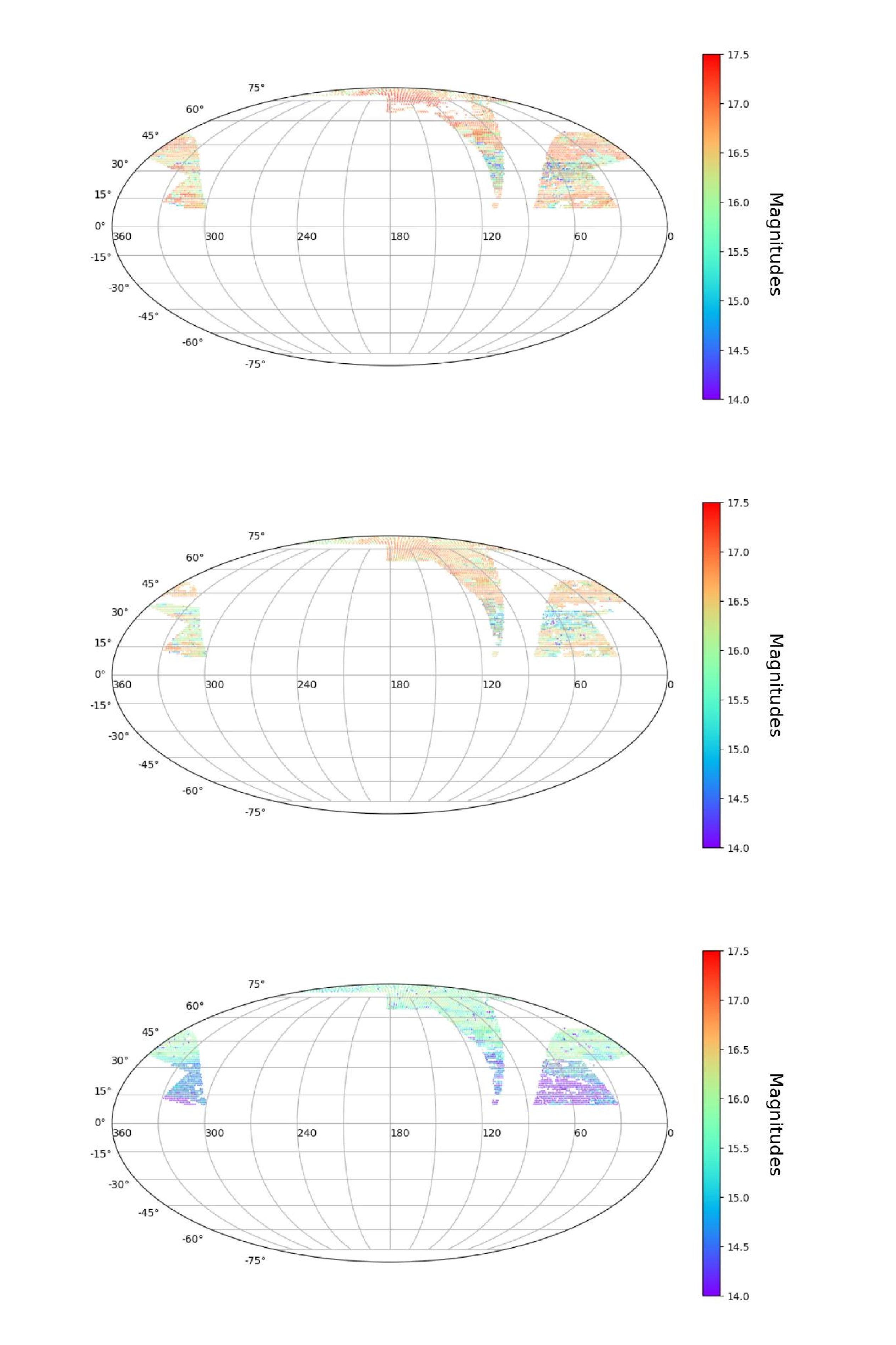}
    \caption{Magnitude distribution for all field of g(upper panel), r(second panel) and r(bottom panel) passbands for S/N=100, corresponding to the photometry error of 0.01 mag. The median magnitude is 16.57, 16.46 and 15.49mag respectively.}
    \label{fig6}
\end{figure}

The limiting magnitudes reach 16.57, 16.46, and 15.49 mag in the g/r/i band, respectively with an uncertainty of 0.01 mag, which is the Possion error, not including the flux calibration errors (internal calibration and zero-point). We use all the frames to calculate the limit mag then use the median value for each band. The magnitudes limit for the uncertainty of 0.2 mag(corresponding to S/N=5) are 20.19, 20.03 and 19.18 respectively.
Figure~\ref{fig6} shows the limit magnitude distribution for all fields of g-band for S/N=100, corresponding to the photometry error of 0.01 mag.

\begin{figure}
    \centering
    \includegraphics[width=10cm]{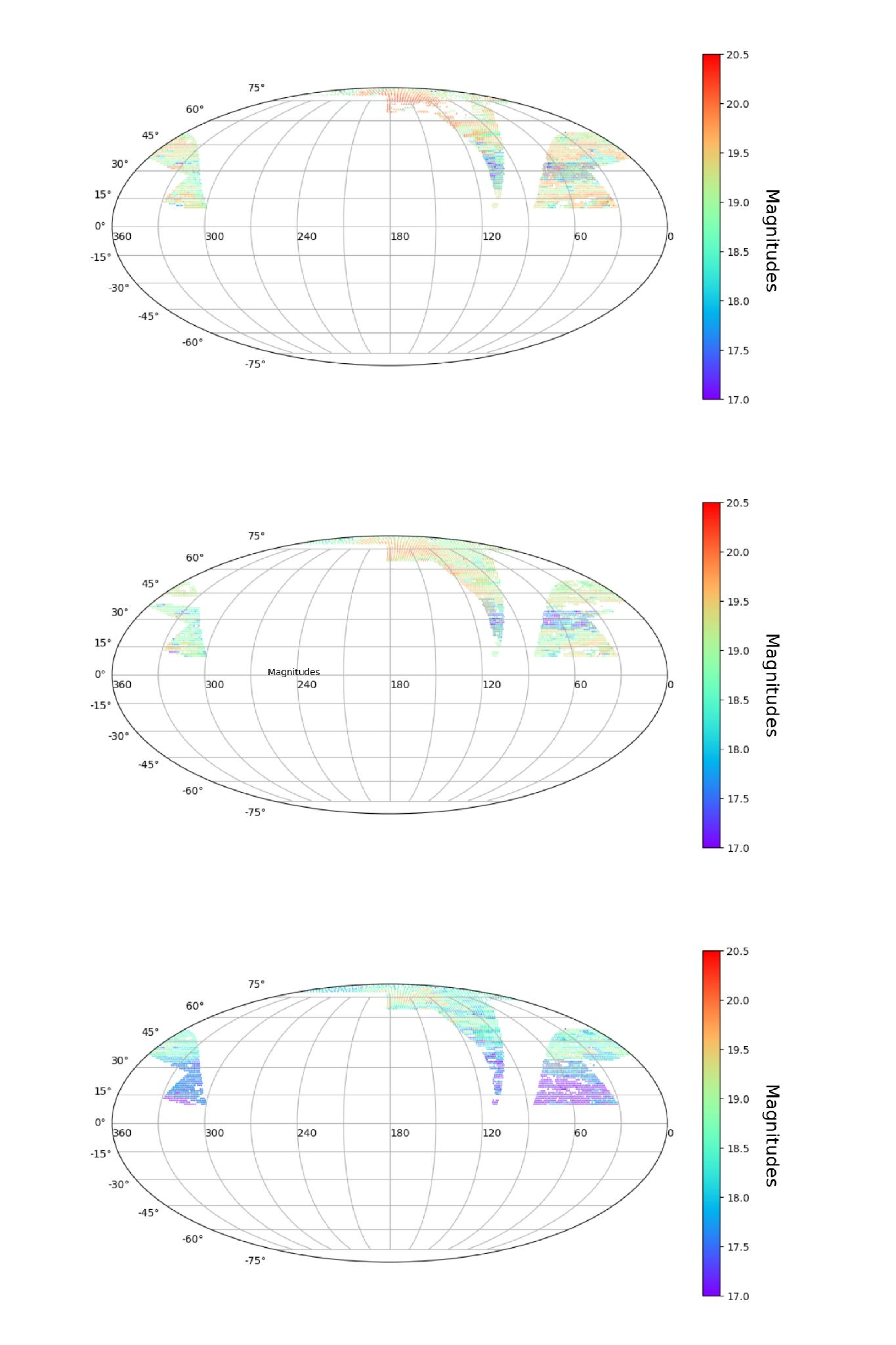}
    \caption{Magnitude distribution of g(upper panel), r(second panel) and r(bottom panel) passbands but for complete magnitudes. The median complete magnitude is 19.2mag, 19.1mag, and 18.2mag, respectively.}
    \label{fig7}
\end{figure}

Figure~\ref{fig7} is the magnitude distribution for g/r/i passbands. From the turning point of the diagrams, the complete magnitudes for g/r/i passbands are 19.2, 19.1, and 18.2 mag respectively, which show the depth of observation.

\begin{figure}
    \centering
    \includegraphics[width=15cm]{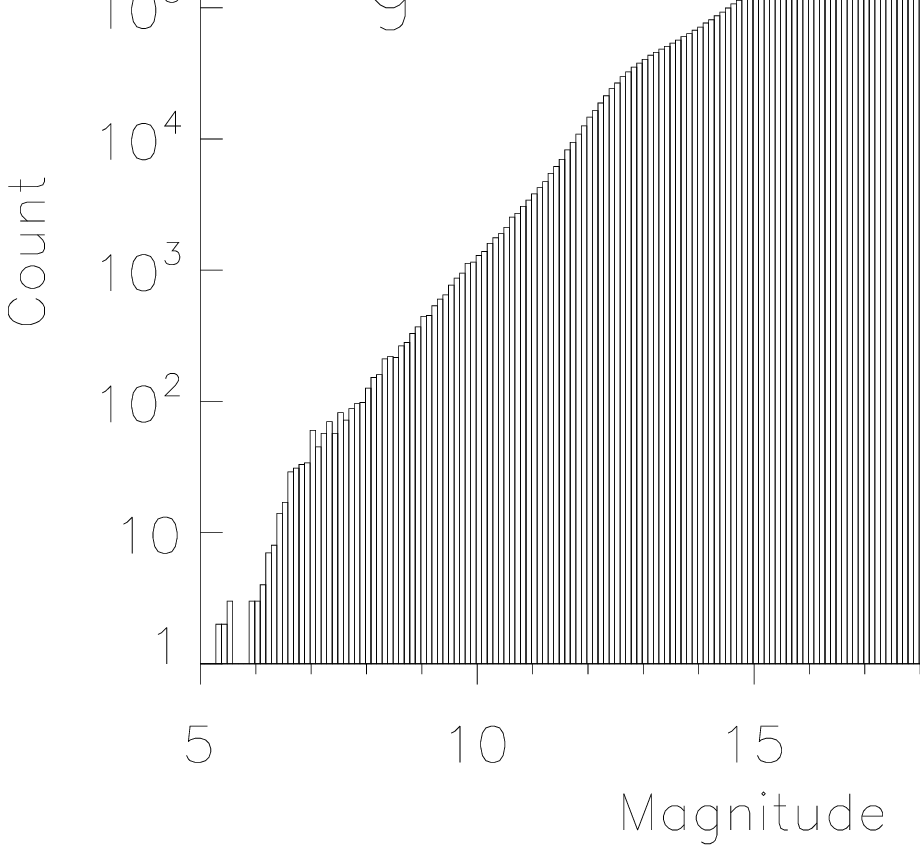}
    \caption{The magnitude distribution for all the sources of the nanshan run of SAGES in all observed fields of g/r/i passbands, respectively. The x-axis is the magnitude range and the y-axis represents the counts for the photometry magnitude number. The peaks are g$\sim$19.2 mag and r$\sim$19.1 mag and i$\sim$18.2 mag.}
    \label{fig8}
\end{figure}

Figure~\ref{fig8} shows the magnitude distribution for all the sources of the Nanshan run of SAGES in all observed fields of g/r/i passbands, respectively. The x-axis is the magnitude range and the y-axis represents the counts for the photometry magnitude number. It can be seen that the peaks are g$\sim$19.2 mag and r$\sim$19.1 mag and i$\sim$18.2 mag for the Nanshan observing run of SAGES, which correspond to the complete magnitude.

\section{Data Access}
SAGES g/r/i data has been released and is available on the China-VO platform and the
National Astronomical Data Center (NADC) https://nadc.china-vo.org.

\section{Summary and Future Plans}
This paper presents part of the whole SAGES, the observing strategy, observation, data reduction, and catalog production of SAGES g/r/i passband witch processed with NOWT.

In the further, some of the data reduction and calibration process steps will be upgraded. For instance, the removal of the fringe of i band images. Hoping this will help to improve photometry and calibration, and provide us with deeper detection and more precise data.

The observation of other passbands of SAGES (DDO51 and $\alpha_w$) will be completed in the next few years. Further works of SAGES will contain more passbands, more images, and better sky coverage.

\normalem
\begin{acknowledgements}

This study is supported by the National Natural Science Foundation of China (NSFC) under grant Nos.12261141689, 12090044, and 12090040. This study is also sponsored by the Xinjiang Uygur Autonomous Region 'Tianchi Talent' Introduction Plan.

The Stellar Abundance and Galactic Evolution Survey (SAGES) is a multi-band photometric project built and managed by the Research Group of the Stellar Abundance and Galactic Evolution of the National Astronomical Observatories, Chinese Academy of Sciences (NAOC).

\end{acknowledgements}

\label{lastpage}
\end{document}